\documentclass[conference]{IEEEtran}
\usepackage{cite}
\usepackage{amsmath,amssymb,amsfonts}
\usepackage{algorithmic}
\usepackage{graphicx}
\usepackage{textcomp}
\usepackage{todonotes}
\usepackage{xcolor}
\def\BibTeX{{\rm B\kern-.05em{\sc i\kern-.025em b}\kern-.08em
    T\kern-.1667em\lower.7ex\hbox{E}\kern-.125emX}}

\begin{document}
%
\title{Death by AI: \\Where Assured Autonomy in Smart Cities Meets the End-to-End Argument}


\author{\IEEEauthorblockN{Gregory Falco}
\IEEEauthorblockA{
Stanford University\\
Stanford, CA USA\\
falco@stanford.edu}
}

\maketitle

\begin{abstract}
A smart city involves critical infrastructure systems that have been digitally enabled. Increasingly, many smart city cyber-physical systems are becoming automated. The extent of automation ranges from basic logic gates to sophisticated, artificial intelligence (AI) that enables fully autonomous systems. Because of modern society's reliance on autonomous systems in smart cities, it is crucial for them to operate in a safe manner; otherwise, it is feasible for these systems to cause considerable physical harm or even death. Because smart cities could involve thousands of autonomous systems operating in concert in densely populated areas, safety assurances are required. Challenges abound to consistently manage the safety of such autonomous systems due to their disparate developers, manufacturers, operators and users. A novel network and a sample of associated network functions for autonomous systems is proposed that aims to provide a baseline of safety for autonomous systems. This is accomplished by establishing a custom-designed network for autonomous systems that is separate from the Internet, and can handle certain functions that enable safety through active networking. Such a network design sits at the margins of the end-to-end principle, which is warranted considering the safety of autonomous systems is at stake as is argued in this paper. Without a scalable safety strategy for autonomous systems as proposed, assured autonomy in smart cities will remain elusive. 

\end{abstract}

\begin{IEEEkeywords}
Smart City, Assured Autonomy, Safety, Autonomous System, End-to-End Argument, Active Networking, Cyber Physical System

\end{IEEEkeywords}

\section{Introduction}
\par Considering the scale at which smart cities will employ autonomous systems, safety assurances will be required. This paper proposes an active network that could facilitate the safe operation of autonomous systems in a smart city ecosystem. The argument for active networking invokes an exception in the end-to-end argument citing how performance enhancements could justify functions at the network level. In this case the performance enhancement is safety, which is a priority considering it can reduce the risk of death caused by an autonomous, cyber-physical system in a densely populated smart city environment. For the sake of spurring discussion, the paper proposes sample functions that could reside at a low level of the stack in order to guarantee certain safety parameters across disparate autonomous systems. 
\subsection{Situation}
As population densities have increased in cities, there has been an incremental toll on urban critical infrastructure's performance. To alleviate this burden by enabling efficiency and cost-savings, there has been movement towards digitizing urban systems by instrumenting critical infrastructure with embedded systems and connecting these systems to the internet. The digital instrumentation and connectivity of industrial systems is referred to as the Industrial Internet of Things (IIoT). The subsequent connectivity of urban systems and services as a result of the deployment of IIoT results in the establishment of a "Smart City". An example of a smart city capability is traffic demand management. To improve the traffic flow during peak demand hours in a city, a combination of traffic sensors, traffic lights, CCTV cameras and traditional traffic police services could be leveraged to minimize congestion, thereby reducing wait times, infrastructure wear and environmental impact (e.g. air pollution). 
\par The instrumentation of urban systems is not new, but the mechanisms for controlling these IIoT systems are changing rapidly. Originally, these IIoT systems may have employed basic ladder logic to facilitate the control of their surroundings. The cost of processors and memory on these IIoT devices has dropped drastically which has enabled many new features for the IIoT - one of which includes the ability to conduct complex processing on the edge. One use case for edge processing is using artificial intelligence (AI) to make control decisions. Such artificial intelligence can vary from enacting AI planning instructions to facilitating distributed deep learning processes. 
\subsection{Complication}
\par Edge AI has accelerated the capabilities of autonomous, cyber-physical systems. Smart cities are increasingly employing this technology for use cases such as autonomous transportation and distributed energy generation and demand management. Because the decision-making for these systems are completed without human oversight, in real-time, there is no room for error. For example, faulty communications between a renewable energy system and a neighborhood's microgrid potentially caused by poor system integration could result in a brown-out. Equally disconcerting is the failure of an autonomous vehicle to perform as intended, resulting in a traffic accident that perhaps causes physical harm or death to riders. 
\par Faults in autonomous systems as described above could be a function of multiple issues. Because of their cyber-physical nature, faults could be caused by mechanical failures. Mechanical issues are largely avoidable through proper commissioning and maintenance plans for the technology. More disconcerting are digitally caused failures such as ones relating to faulty AI logic or malicious actors via security vulnerabilities. These digital faults are problematic because they are difficult to troubleshoot, could remain undetected in the system for a prolonged period and could cause catastrophic failures in the autonomous system. Further, such digital issues could cause cascading failures across multiple autonomous systems that have logic that rely on each other - as is the case for renewable energy systems.
\par Perhaps the most important distinction about autonomous systems in smart cities is that because of their cyber-physical nature, a premium is placed on the physical safety of the agent. It is critical for the control of these systems to be consistent and reliable in order to facilitate safety. Without safety as a priority, these systems could ostensibly result in human harm or death due to their cyber-physical operations. This is especially important in smart cities where hundreds or even thousands of autonomous systems are operating and potentially interacting amidst densely populated urban centers. This begs the question: how do we instill safety features in every autonomous system across a given operating landscape or ecosystem? Because of the diverse field of developers, manufacturers, operators and users of such autonomous systems, these safety parameters should be instantiated in the common element across the agents - the network. 
\subsection{Proposal}
\par The proposed Assured Autonomous Cyber-Physical Ecosystem (AACE) network architecture will draw from other custom, use case-specific networks that employ active networking. The ambition of AACE is to establish a baseline of safety across all autonomous systems that reside on the network. Such a baseline would enable public trust in the safety of cyber-physical, autonomous systems that operate at scale across a smart city. 
\section{Prior Art}
Since the ubiquity of the Internet, there have been calls from industry groups and various governments and their constituent agencies for custom, use case specific networks that stand alone from the global Internet. Generally, these calls have been the result of specific needs that the Internet does not fully address. This has taken various forms. For example, the space industry is currently designing the "space internet" to facilitate communications across satellite systems in space \cite{spaceinternet}. Similarly, the government of Russia has set out to build its own "unplugged", sovereign internet \cite{russiainternet}. The ambition with such networks is that they would be developed with a particular user group in mind and interact with the global Internet in discreet ways - if at all. Nearly all instances are a reflection of the need for greater application control within a network. 
\subsection{Network Architecture Design}
\par The internet was designed using a set of design principles called the end-to-end arguments \cite{blumenthal2001rethinking}. The end-to-end arguments dictate that "functions placed at low levels of a system may be redundant or of little value when compared with the cost of providing them at that low level" \cite{saltzer1984end}. The principle continues to dictate that providing application functionality in the network allocates unnecessary strain to all applications that interact with the network - whether they make use of the application functionality or not. It also argues that because the network may not have as much information as the application, the low level subsystem cannot perform as efficiently compared with if similar requirements were in the application layer \cite{saltzer1984end}. 
\par Saltzer, Reed and Clark, the authors of the end-to-end argument, permit that there are exceptions to this rule of thumb. They assert that low level functionality could become justified as a performance enhancement \cite{saltzer1984end}. A major consideration is if the performance enhancement afforded by functionality at the low level is worthwhile to all the devices that reside on the network, despite additional overhead caused by the function. This question has been grappled with by scholars discussing the merits of active networking.
\par Bhattacharjee, Calvert and Segura describe an extension of the end-to-end argument called "active networking" which rationalizes functionality in the network. They propose a performance model to assess the benefits of functions residing in-network versus in the application \cite{bhattacharjee1997active, bhattacharjee1997architecture}. They argue that placing higher level functionality in the low levels of the network is not contradictory to the end-to-end principle, rather the principle merely qualifies a notional threshold of trade-offs be met before designing functions in the network. For example, there could be use cases where certain environmental information is held within the network and timely use of such information could enhance the services of the application. They proceed to propose some functions that could be of benefit across a class of systems on a network including: time and place of congestion, global patterns across the network, and the location of packet losses \cite{bhattacharjee1997active}. These classes of functions are metadata that could have direct benefit to application functionality. To serve such functions, it would be critical that all applications could receive a benefit from these functions in spite of any overhead required to serve them. The team proceeded to calculate, using their model, a cost-benefit of employing functions on the network versus the application layer for reliable multicast and application-specific congestion control. In both cases, the team determined that employing active networks outperformed the end-to-end network implementation where functions reside on the application \cite{bhattacharjee1997active}. 
\par Research in active networks peaked in the mid 1990's amidst the DARPA Working Group on Active Networks \cite{calvert1999architectural}. An insight from Calvert, one of the participants of the DARPA program, was that it is difficult to quantify benefits of active networking; yet the benefit of improved manageability and control of the network is widely cited \cite{calvert2006reflections}. The notion of active networking improving the control of applications is also reflected in a survey on active networking research \cite{tennenhouse1997survey}. The survey team concludes there is a major mismatch between the rate user requirements change and the status of network infrastructure, which can be interpreted as a control issue. They suggest that the future will entail protocol components instead of stacks which are highly specialized to the features of a given use case. 
\subsection{Active Networking in Action}
\par Despite the design thinking behind active networks being discussed in the 1990s, there seems to be little discussion networks being implemented in this manner until much later. Control continues to be a core theme for those that seek active networks. For example, Russia's quest to build the RUNET is largely a function of their interest in having complete control over their segment of the Internet. Their interest in control is in the name of national security \cite{nikkarila2017runet}. The RUNET has been discussed for over a decade, but only recently has this active network been tested "successfully" where Russia claims to have disconnected itself from the Internet on December 23, 2019 \cite{tucker}. As part of this, specific control features were tested such as “the possibility of intercepting subscriber traffic and revealing information about the subscriber, blocking communication services" according to a state media document published on December 5, 2019 \cite{tucker}. Russia is not the first country to work towards controlling their citizens' interactions with the Internet. Both Iran and China also have versions of their own active networks to enforce control over their citizens \cite{wakefield}.
\par Active networks have also been discussed for specific industry use cases in recent years. For example, the space internet, while in many ways aims to mirror a "horizontal" infrastructure akin to the global Internet, intends to have various architectural designs to serve different purposes - namely, Earth Science Enterprise, Human Exploration and Development in Space, and Space Science Enterprise \cite{bhasin2001space}. The variations of architecture to serve these different purposes are to facilitate a customized level of application control for their given use case. 
\par Over the past decade, the energy industry has also called for their own network for energy distribution. In these cases scholars have designed energy routers to facilitate the flow and delivery of energy through distributed means \cite{xu2011energy}. Control issues are cited as a barrier to the implementation of distributed generation \cite{sun2015multiagent}. Considering the importance of energy distribution as critical infrastructure, it is essential for operators to have control over the applications that facilitate energy delivery. 
\subsection{Safe Cyber-Physical Systems}
\par Control is a major theme across network architectures that present exceptions to the end-to-end principle. Saltzer, Reed and Clark posited that low level network functionality could be justified for performance enhancement, which may well include improved control. Considering the safety of autonomous, cyber-physical systems is primarily a control issue, active networking could prove useful to facilitate safety of these systems. Further, Bhattacharjee, Calvert and Segura cost/benefit calculations for active networking suggested that functions that help to orient an application in relation to its environment (such as the location of congestion) is better situated in the network than in the application \cite{bhattacharjee1997active}. Autonomous cyber-physical systems' safe use is generally reliant on the system's ability to orient itself to its surroundings, which also points towards the use of active networking to enable safety of these systems. Perhaps the most compelling reason to consider the exception to the end-to-end principle for an autonomous systems network is because low level functionality enables safety at the expense of performance, which is certainly justifiable when lives may be at stake. 
\par Leveson, a long-established expert in safe software systems, asserts that safety issues arise because of the interactions across "unfailed" components \cite{leveson2020you}. She describes how systems cannot be modeled or analyzed for safety in isolation of other interacting systems and that testing, simulation and standard formal verification is ineffective at assessing a system's safety. While such tests may be ineffective to assess safety on individual systems, perhaps safety is better assessed at the system ecosystem or network level. This could especially be worthwhile to assess if there is in fact active networking in place to ensure the interactions across systems are safe. In the subsequent section, we propose functionality that belongs in the autonomous system's network to enable safety. 
\section{Architectural Design Considerations}
\par Rather than the applications themselves be designed for safety, it will be important to provide guarantees that the entire ecosystem of applications is indeed safe by embedding safety parameters in the network. Employing active networking would help to facilitate this on a specialized AACE. Such an architectural design is uniquely applicable to when there are many autonomous systems deployed at once. A smart city is an example for when a diverse group of autonomous systems will perform at scale. A rationale for separating autonomous systems from the Internet is provided below. Then, sample functions and their associated concepts for AACE are proposed that could enable the safety of autonomous systems at scale. At this time, it is unclear if these are the right functions to include in an active network. It is also not yet known what the overhead cost and relative benefit for each of these sample functions may be. These sample functions are described primarily for the sake of discussion. Functions that will be discussed include: demand management, proximal insight, reliable multicast, conflict detection and controlled failure. A rationale for each function's inclusion is included below. 
\subsection{A Separate Network}
\par There are three principal reasons why a separate network for AACE is warranted. First, air-gapping autonomous systems from the global Internet could increase the work-factor of bad actors seeking to cause harm to these systems. While not foolproof by any means as demonstrated in previous successful attacks \cite{zetter2014countdown}, air-gapping critical functions could increase the work-factor of a would-be attacker that intends to corrupt the safety of an autonomous system. Second, certain cyber-physical systems are very sensitive to errant network activity because they are highly calibrated and/or delicate. For example, there are anecdotal reports that systems like supervisory control and data acquisition (SCADA) controllers (employed for long-range industrial control) break during penetration testing or high network activity periods. Considering autonomous systems would also be highly calibrated considering their sophistication, it is feasible for errant network activity over the Internet to cause disruptions, thereby compromising system safety. Finally, establishing a new and separate network for autonomous systems could provide the active networking functions described below.
\subsection{Demand Management}
\par A given smart city will ostensibly have hundreds of thousands, if not millions of sensors and associated controllers across the urban environment. Assuming each cyber-physical, autonomous system has it's own intranet or communication system like a CANbus and the IIoT components of a given autonomous system interact with other autonomous systems in a unified way, there are still likely to be thousands of autonomous systems that need to interact across an ecosystem at any time. Because of this scale of operations, it will be important for the systems to have realistic expectations of communication latency, such that it can modulate its decision-making speed accordingly. 
\par For example, an autonomous vehicle may assume that it will receive information from a dog collar about its physical location in the middle of the road during a green light in real-time. The autonomous vehicle may allow for a 2-second reaction time window to receive new emergency location data so that it can act on this information. However, in this hypothetical situation, there are so many devices on the network at a certain time that there is a 3-second delay across the network from when the dog collar's location state is captured to when it is received by the vehicle. In such a scenario, it would be critical for the vehicle to know about this delay and account for its allowable reaction-time window accordingly (by slowing down). If the network is equipped with a demand management function like this, autonomous systems could be better equipped to handle situations that could compromise safety due to latency. 
\subsection{Proximal Insight}
\par Because all systems will need to interact with each other in some capacity, each should be aware of the state of the most proximal autonomous systems at any given time. More specifically, the network should be able to serve each system with information regarding the latest request of proximal systems. Such requests could provide insight to changes in physical course, data load requirements of the network or emergency signals being emitted by local systems. 
\par For example, it could be highly valuable for a network to consistently provide all autonomous systems with data concerning other systems in distress. This could provide both a mechanism that supports autonomous system resilience facilitated by the network and a mechanism for warnings about immediate safety threats to given autonomous systems that are part of the ecosystem.
\subsection{Reliable Multicast}
\par Reliable delivery of data to all necessary systems will be crucial for the safety of autonomous devices. Generally, data loss is common and accepted across many networking protocols. This is especially true for point to multi-point data delivery. The safety of autonomous systems will be reliant on state communication from other autonomous systems so that any given system can orient itself in relation to the rest of the ecosystem. For this reason, it will be unacceptable for data to not reach any node across the network. 
\par For example, if a drone plans to land in a given location, to avoid collisions, it will be critical to deliver its coordinates out to other systems. Should only a limited number of systems receive the notification of landing from the drone, there is an increase in the likelihood that a collision occurs. Therefore, systems need to be assured that all applications will reliably receive data delivered. If data is not reliable received, systems should be notified accordingly so that they can make decisions about how to handle the communication failure accordingly.
\subsection{Conflict Detection}
\par Considering the scale at which autonomous systems will operate in a smart city, it is inevitable that conflicting commands will be issued to various systems which may lead to a safety issue. Rather than rely on the application to catch discrepancies across commands, it would be valuable for meta-level conflict detection to occur on the network. The network could help flag conflicting information for the systems and provide validation to the command as unique, rather than waiting until the conflicting logic is processed in the application. 
\par For example, imagine a scenario where a smart parking application dictates that the same space is available to two different vehicles at the same time. Both vehicles then proceed to take action towards securing the open spot. It would be valuable for the network to understand the constraints of the system environment and by using planning logic conclude that it is impossible for both vehicles to park in the same spot despite both taking actions to secure the space. The network could proceed to signal to one of the vehicles that a conflict was detected and therefore the spot is no longer available. Such an action may not only spare the possibility of a collision.
\subsection{Controlled Failure}
\par Perhaps one of the most useful safety features of an active network for a smart city ecosystem would be to enable safe failures of the autonomous systems on the network. There could be infinite possibilities for the interaction of systems to cause unsafe conditions. Rather than having a network that can protect against all of these potential safety hazards, it would be valuable for the network to allow for the safe and controlled failure of autonomous systems. One way that a network could perhaps do this is through a distributed voting mechanism where each application is requested by the network to vote on the safety of each system present. Should all nodes return with consensus regarding a bad actor node, the network would be able to eject the "bad actor" from the network in a controlled way. A mechanism as described would help to purge the network from potentially hazardous agents and continuously prune the network. When the network determines a node should be decommissioned, it is first quarantined and allowed to complete its most recent command as long as it does not impact other nodes on the network. The application is simultaneously notified by the network that it will be cut off and should plan for a safe dismount accordingly. Equally, any autonomous system that was communicating with the bad actor would also be notified that the bad actor will be decommissioned. 
\par An example of this could involve a smart street light that is continuously flashing and simultaneously pinging nearby autonomous systems about their error unnecessarily, causing packet flooding across the network. The street light could be voted out of the network by the other nodes. Th network would first notify the smart street light of the issue as well as notify all systems that are communicating with the street light and vice versa that the system's communications will be decommissioned. Then the network will cease communication of signals that the light is disseminating to other systems.  Establishing an integrity check through a voting mechanism would provide increased reliability and trust across the network. 
\section{Discussion}
\par Smart cities will ultimately rely on autonomous systems to facilitate critical infrastructure operations, which comes with considerable benefits. Cyber-physical, autonomous systems could yield economic efficiencies, reduce environmental impact and increase convenience for citizens. However, safety risks are sure to accompany these autonomous systems. 
\par Safety has been addressed for autonomous systems before by establishing processes for ensuring safety in systems at the software or process level \cite{leveson1995safeware,ishimatsu2010modeling,friedberg2017stpa}. While these means for facilitating safety could be highly effective for individual systems, smart cities will realize a scale of operations of autonomous systems not previously expected. This warrants a safety enforcement mechanism that could be consistent across all systems regardless of their function or how they were developed. Employing network-level safety functions will ensure that there is a scalable way to implement a baseline level of safety across all autonomous systems. There are likely to be cost trade-offs with employing functions at this low level, but this is justified by the universal enhancement of safety and improved control over the otherwise potentially dangerous systems.
\section{Conclusion and Future Work}
\par This paper intends to provide a starting point for discussion about active networking for autonomous systems and functions that should be built into AACE. Considerable research is still needed to determine which are the most critical functions required at the network level of a smart city ecosystem to ensure safety across all autonomous devices. Then, evaluating the technical feasibility of implementing these functions on the network level is required. Also, a cost/benefit calculation should be completed to evaluate if the function should indeed be at a low level of the stack. 
\par As autonomous systems become ubiquitous and their operations scale accordingly, there will be a vital need for reliable control to ensure safety and reduce the risk of death by AI-enabled autonomous, cyber-physical systems. Not only must these systems operate in a safe way themselves, but their interactions with other systems must also be assuredly safe at scale. Therefore, establishing safety controls at the lowest common denominator across all interacting autonomous systems - the network, could provide considerable benefits given the diversity of systems, developers, manufacturers, operators and users that could otherwise lead to safety failures.  Creating safety assurances by embedding certain functions in the network can take us a step closer to legions of safe, autonomous cyber-physical systems for future smart cities.

\bibliographystyle{./bibliography/IEEEtran}
\bibliography{./bibliography/IEEEabrv,./bibliography/IEEEexample}

\end{document}